\documentclass[journal]{IEEEtran}
\usepackage[T1]{fontenc}
\usepackage[utf8]{inputenc}
\usepackage{amsmath,amsfonts}
\usepackage{algorithmic}
\usepackage{algorithm}
\usepackage{array}
\usepackage[caption=false,font=normalsize,labelfont=sf,textfont=sf]{subfig}
\usepackage{textcomp}
\usepackage{stfloats}
\usepackage{url}
\usepackage{verbatim}
\usepackage{graphicx}
\usepackage{cite}
\usepackage{afterpage}
\usepackage{placeins}
\usepackage{makecell}
\usepackage{graphicx}
\usepackage{multirow}
\usepackage{amssymb}
\usepackage{float}
\usepackage[breaklinks]{hyperref}
\usepackage[table,xcdraw]{xcolor}
\usepackage{tabularx}
\usepackage{ulem}
\renewcommand{\arraystretch}{1.5} % 设置行高
\begin{document}

\title{ATHENA: An In-vehicle CAN Intrusion Detection Framework Based on Physical Characteristics of Vehicle Systems}

\author{
Kai Wang, \IEEEmembership{Member,~IEEE}, 
Zhen Sun, 
Bailing Wang, \IEEEmembership{Member,~IEEE}, 
Qilin Fan, \IEEEmembership{Member,~IEEE},

Ming Li, \IEEEmembership{Member,~IEEE}, 
Hongke Zhang, \IEEEmembership{Fellow,~IEEE}

\thanks{This work is supported by National Natural Science Foundation of China (NSFC) (grant number 62272129) and Taishan Scholar Foundation of Shandong Province (grant number tsqn202408112).}

\thanks{Kai Wang, Zhen Sun and Bailing Wang are with School of Computer Science and Technology, Harbin Institute of Technology, Weihai, China, and also with Shandong Key Laboratory of Industrial Network Security, China (e-mail: dr.wangkai@hit.edu.cn; 23s130402@stu.hit.edu.cn; wbl@hit.edu.cn).}
\thanks{Qilin Fan is with School of Big Data and Software Engineering, Chongqing University, Chongqing, China, and also with Key Laboratory of Dependable Service Computing in Cyber Physical Society of Ministry of Education, Chongqing University, Chongqing, China (email: fanqilin@cqu.edu.cn)}
\thanks{Ming Li is with Jinan Key Laboratory of Distributed Databases, Shandong Inspur Database Technology Co., Ltd, Jinan, China (liming2017@inspur.com).}
\thanks{Hongke Zhang is with School of Electronic and Information Engineering, Beijing Jiaotong University, Beijing, China (e-mail: hkzhang@bjtu.edu.cn).}

\thanks{Corresponding author: Kai Wang and Bailing Wang.}
}

\maketitle
\begin{abstract}
With the growing interconnection between In-Vehicle Networks (IVNs) and external environments, intelligent vehicles are increasingly vulnerable to sophisticated external network attacks. 
%The current IVN intrusion detection methods ignore the vehicle-cloud integrated architecture and the physical characteristics of the vehicle system, and fail to fully consider the key requirements of vehicle-cloud coordination and real-time performance.
%In this paper, we propose ATHENA, the first IVN intrusion detection framework using vehicle-cloud integrated architecture that exploits the first-principled physical knowledge of the vehicle system to efficiently improve the quality of intrusion detection rule mining, and achieves resource-friendly coordination between the vehicle side and the cloud side that balances task performance and resource constraints. 
%This paper proposes ATHENA, the first IVN intrusion detection framework that adopts a vehicle-cloud integrated architecture to achieve an acceptable balance between task performance and resource constraints. 
This paper proposes ATHENA, the first IVN intrusion detection framework that adopts a vehicle-cloud integrated architecture to achieve better security performance for the resource-constrained vehicular environment. 
Specifically, in the cloud with sufficient resources, ATHENA uses the clustering method of multi-distribution mixture model combined with deep data mining technology to generate the raw Payload Rule Bank of IVN CAN messages, and then improves the rule quality with the help of exploitation on the first-principled physical knowledge of the vehicle system, after which the payload rules are periodically sent to the vehicle terminal. 
At the vehicle terminal, a simple LSTM component is used to generate the Time Rule Bank representing the long-term time series dependencies and the periodic characteristics of CAN messages, but not for any detection tasks as in traditional usage scenarios, where only the generated time rules are the candidates for further IVN intrusion detection tasks.
Based on both the payload and time rules generated from cloud and vehicle terminal, ATHENA can achieve efficient intrusion detection capability by simple rule-base matching operations, rather than using complex black-box reasoning of resource-intensive neural network models, which is in fact only used for rule logic generation phase instead of the actual intrusion detection phase in our framework.
%Through the protection of these two dimensions, the resources on the cloud and vehicle side can be allocated more optimally, thereby improving the performance of intrusion detection in vehicle networks. 
Comparative experimental results on the ROAD dataset, which is current the most outstanding real-world in-vehicle CAN dataset covering new instances of sophisticated and stealthy masquerade attacks, demonstrate ATHENA significantly outperforms the state-of-the-art IVN intrusion detection methods in detecting complex attacks. We make the code available at \url{https://github.com/wangkai-tech23/ATHENA}.
\end{abstract}


\section{Introduction}

\IEEEPARstart{I}{n-Vehicle} Networks (IVNs) play a crucial role by enabling real-time connectivity in modern vehicles, supporting advanced features such as autonomous driving, in-car entertainment, smart navigation, and remote diagnostics\cite{A1,A57}. Through these networks, vehicles can interact with other cars, road infrastructure, cloud platforms, and smart devices, significantly enhancing driving safety, comfort, and traffic efficiency. However, as the degree of network integration in vehicles increases, their digital platforms become exposed to more complex network environments. This makes vehicles vulnerable to various cyber threats, such as hacking, data breaches, and remote control attacks, posing significant security risks for both car owners and manufacturers\cite{A2,A58}.

Modern intelligent vehicles widely adopt steer-by-wire systems, relying on Electronic Control Units (ECUs) to transmit data via the Controller Area Network (CAN)\cite{A3,A4,A5,A6}. The payload part of CAN data has 8 bytes to carry the physical attribute information according to the definition of CAN communication matrix of automobile manufacturers. Additionally, the CAN is exposed through the OBD-II port and mobile networking interfaces, making it susceptible to attacks. Therefore, research on CAN vulnerabilities and Intrusion Detection Systems (IDS) has been steadily increasing\cite{A7,A8}.

\begin{figure}[!ht]
    \centering
    \includegraphics[height=3.0cm]{"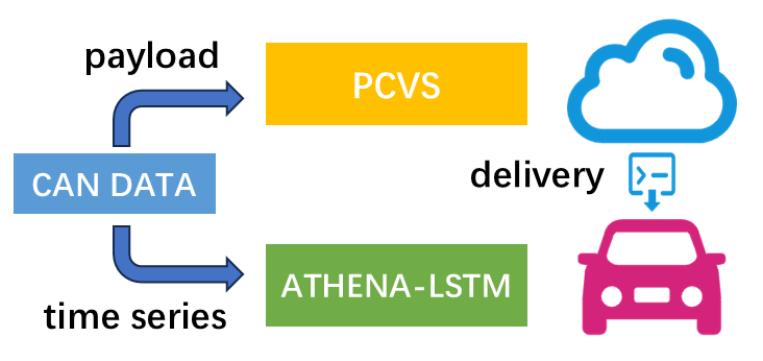"}
    \caption{Simple ATHENA framework.}
    \label{fig_1}
\end{figure}

In light of this, existing research has adopted various methods to detect intrusions targeting the CAN bus\cite{A35,A36}. The first approach is Rule/ Specification-Based methods\cite{A13}. While these methods offer high detection efficiency, the rules are generated based on human expertise, which can not fully represent the operational patterns of the system. Frequency/Timing-Based methods are primarily used to detect Timing Transparent attacks, such as DoS attacks, fuzzing attacks, etc.\cite{A63,A64}, but they are unable to identify masquerade attacks. Although Machine Learning-Based methods perform well, model training is time-consuming, and the data used for training do not fully consider the physical attribute signals carried by CAN message payloads, and the current research mainly focuses on vehicle detection\cite{A28,A64,A49}. However, the computational resources on the vehicle end are typically limited, and the balance of computational power within the vehicle-cloud integrated architecture has not been adequately considered.

To solve the above problems, we propose ATHENA framework, which constructs a new resource allocation architecture based on the integration of vehicle and cloud architecture by deploying the massive rule generation tasks guided by the physical characteristics of the vehicle in the cloud, and the lightweight time series feature mining model training and rule-based reasoning tasks in the vehicle. The ATHENA framework consists of two modules: the Vehicle-Cloud Processing and Verification Module (PCVS) and the ATHENA-LSTM Module. The PCVS Module integrates the first-principle physical knowledge of vehicle systems into data-driven rule mining to address complex masquerade attacks. The cloud performs computationally intensive data mining to guide the generation of the Payload Rule Bank through the physical characteristics contained in CAN messages, which exhibits high robustness and effectively defends against advanced masquerade attacks. The Payload Rule Bank only needs to be periodically updated and sent to the vehicle for fast inference and detection. In contrast, ATHENA-LSTM Module captures the long-term dependencies in the transmission cycles of CAN messages, performing training and detection on the vehicle terminal to respond in real time to low-level attacks triggered by periodic changes. ATHENA achieves a very reasonable balance between high computing power and low resource consumption by deploying computation-intensive tasks to generate the Payload Rule Bank in the cloud, mining temporal rules to generate the Time Rule Bank on the vehicle terminal, and performing rule-based inference detection according to the two rule banks on the vehicle terminal.

By synergizing these two modules, the ATHENA framework not only significantly enhances the security of in-vehicle networks but also fully utilizes the computational capabilities and real-time responsiveness of the cloud-vehicle integrated architecture, providing a ground breaking paradigm for intrusion detection in the future Internet of Vehicles (IoV).

In summary, our main contributions are as follows:

\begin{itemize}
\item We propose ATHENA framework, which is the first to utilize the bit-level physical characteristics of CAN messages of vehicle systems to guide the mining of invariant rules and the training of model. According to the different dimensions of CAN message detection and the different consumption of computing power resources, the training and detection of PCVS and ATHENA-LSTM modules are deployed in the cloud and the vehicle. In this way, the cloud can complete the massive rule generation task and build the physical rule bank of the vehicle system, and the vehicle terminal is responsible for the lightweight model training and rapid rule-based detection, so as to construct a new resource allocation architecture based on the vehicle-cloud integration architecture.

\item The PCVS module uniquely combines the first-principle physics knowledge contained in the CAN messages of the vehicle system with the data-driven rule mining technology to generate the Payload Rule Bank that reflects the running characteristics of the vehicle system, overcomes the limitations of the traditional engineer defined rule banks, and improves the accuracy and efficiency of attack detection. Moreover, relying on the vehicle-cloud integrated architecture, the PCVS module deploys the massive rule generation task in the cloud, and the rapid rule-based reasoning task in the vehicle, which optimizes the resource allocation of the vehicle-cloud integrated architecture and achieves the best balance between high computing power and low resource consumption.
    
\item ATHENA-LSTM module innovates by learning the "normal sending pattern" of each CAN message ID to replace the traditional LSTM to predict the future output according to the data at the historical moment and its timing pattern, avoiding the misjudgment of the traditional LSTM in in IVNs. Firstly, it calculated and normalized the message sending interval to enhance the processing ability of short period messages, and then used the prediction results and standard deviation to dynamically adjust the temporal rule base to adapt to the changes of different data distribution. In addition, ATHENA-LSTM combines the physical characteristics of the vehicle system and performs rule-based intrusion detection according to the Time Rule Bank, effectively improves the detection efficiency of the Timing Transparent attack, and has higher flexibility, adaptability and reliability.

\item To evaluate the performance of ATHENA framework on the real vehicle-cloud integrated architecture, we propose an innovative experimental design, which uses the in-vehicle computing platform (NVIDIA Jetson Nano T206 with ARM architecture) and the mobile cloud (ROG Strix G814JV\_G814JV with a 13th Gen Intel(R) Core(TM) i9-13980HX @ 2.20GHz GPU and an 8 GB NVIDIA GeForce RTX 4060 GPU) to build the experimental environment. The computationally intensive rule generation task is performed in the cloud, and the rule-based real-time detection is performed autonomously in the vehicle terminal. This novel experimental setup fully leverages the limitations of vehicle computing resources while demonstrating the ATHENA framework's efficiency and scalability in real-world vehicle environments, validating its effectiveness in complex attack scenarios.
\end{itemize}

The reminder of this papper is organized as follows: Section \ref{section3} introduces the attacks on CAN bus and summarizes existing representative IVN intrusion detection methods. Section \ref{section4} describes our proposed ATHENA. We show the experiment setup and results in section \ref{section5} followed by the conclusion in section \ref{section6}. 
\section{Background}\label{section3}%Approach / System Design
In this section, we introduce the attacks on CAN bus in \S \ref{sec2.1} and the related work in \S \ref{sec2.2}.

\subsection{Attacks on CAN bus}\label{sec2.1}

The CAN protocol is widely used in automotive systems due to its real-time performance, reliability, and cost-effective decentralized design. However, its early design does not account for modern security needs, lacking encryption or authentication features for data packets. As IVNs become increasingly integrated with external networks, the CAN bus is increasingly vulnerable to external intrusions\cite{A50,A51,A52,A53,A55,A56}. Recently, several attacks on the CAN bus have occurred, some via wireless network intrusions (e.g., Bluetooth, broadcast channels, addressable channels) and others requiring physical access (e.g., OBD-II diagnostic ports, USB ports). Based on the attack methods, CAN bus attacks can be categorized into three types\cite{A3}:

\textit{1) Fabrication Attacks:}
ECUs process only the last data frame for a given ID.  Fabrication attacks exploit this by injecting malicious messages.  DoS attacks flood with ID 0x000, blocking communication, fuzzing disrupts operations with random messages, and targeted ID attacks compromise critical ECUs with crafted payloads.

\textit{2) Suspension Attacks:}
The attacker who initiates the suspension attack generally only needs to get a weak damaged ECU on the car system (he can only pause the ECU to send messages but cannot control the ECU to send specific messages), and destroys the normal operation of the car physical system by silencing the ECU.

\textit{3) Masquerade Attacks:}
Masquerade attacks are highly advanced and involve long-term planning. The attacker first halts message transmission from a weakly compromised ECU with a specific ID, then injects spoofed messages with the same ID and frequency using a strongly compromised ECU, impersonating the target ECU. Charlie Miller and Chris Valasek\cite{A40}, renowned white hat hackers in automotive security, remotely controlled a Jeep Cherokee using such an attack, leading Chrysler to upgrade its systems and recall 1.4 million vehicles. Masquerade Attacks can have serious consequences, however, there has been a lack of high quality datasets for research until the ROAD dataset proposed by Oak Ridge National Laboratory, USA is available.

Due to the uniqueness of CAN messages in the physical characteristics of the vehicle system, which have different sending cycles according to different IDs, the attacks can be divided into two categories\cite{A2}:

\textbf{Timing Transparent vs. Timing Opaque}

Timing transparent attacks can be detected using frequency-based methods to identify messages that violate the sending period. However, CAN messages have varying sending periods based on different IDs and triggering events, making it crucial to determine the normal sending period interval for attack detection.

Timing opaque attacks, like Masquerade Attacks, maintain CAN message periodicity and evade frequency-based IDS detection. A more advanced approach targeting the message payload is needed.

\begin{figure}[!ht]
    \centering
    \includegraphics[height=2.25cm]{"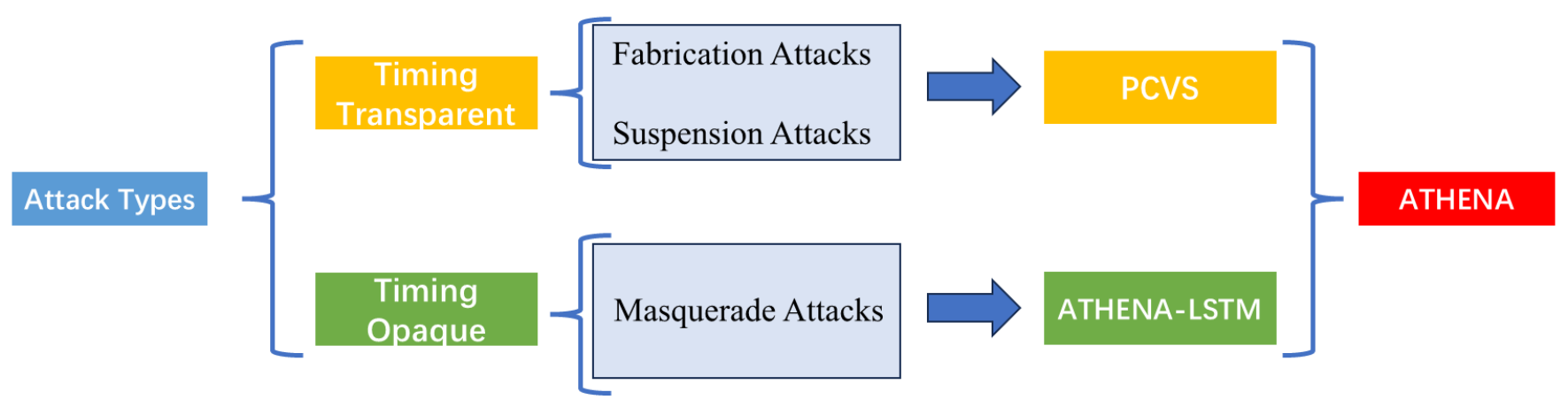"}
    \caption{Attack Types.}
    \label{fig_2}
\end{figure}

In summary, we design the detection idea of our ATHENA framework from these two classifications of different dimensions of attacks, as shown in Figure \ref{fig_2}.

\subsection{Related Work}\label{sec2.2}
fRule/Specification-Based CAN IDS offer extremely fast detection of attacks\cite{A13}. However, the construction of their rule sets and whitelists/blacklists often relies on the experience of engineers, which may not fully represent the actual operational patterns of the physical system. Nevertheless, Cheng Feng et al.'s research\cite{A17} provides us with a new direction of rule generation, through the method of data mining to mine the rules of the actual physical system, so that the generation of rules is more comprehensive. We introduce it into the field of vehicular network intrusion detection for the first time, combining the physical characteristics of the vehicle system and the vehicle-cloud integrated architecture to make the adaptation. Frequency/Timing-Based CAN IDS focus on the periodicity of CAN messages and can effectively detect Timing Transparent attacks\cite{A63,A64}. These systems provide high detection efficiency, smaller model sizes, lower computational resource requirements, and shorter training times, making them suitable for training and detection on resource-limited vehicle end. However, such methods have almost no capability to detect masquerade attacks, particularly those with Timing Opaque, which are highly covert and destructive. Machine Learning-Based CAN IDS are currently the mainstream research direction and hold significant potential in the field of intrusion detection\cite{A28,A64,A49}. However, efficient machine learning models generally have large model sizes, long training times, and substantial computational demands, making it difficult to perform training on resource-constrained vehicle end. Furthermore, existing machine learning-based CAN IDS typically focus on vehicle end, neglecting the vehicle-cloud integrated architecture. As a result, although these methods perform excellently on experimental environment, they struggle to be effectively deployed in real-world environments.

\section{ATHENA Framework}\label{section4}%Approach / System Design

In this section, we give the framework overview of the ATHENA, and then present the details of its key components.

\subsection{Framework Overview}\label{sec4.1}
\begin{figure*}[htbp]
    \centering
    \includegraphics[width=0.95\textwidth]{"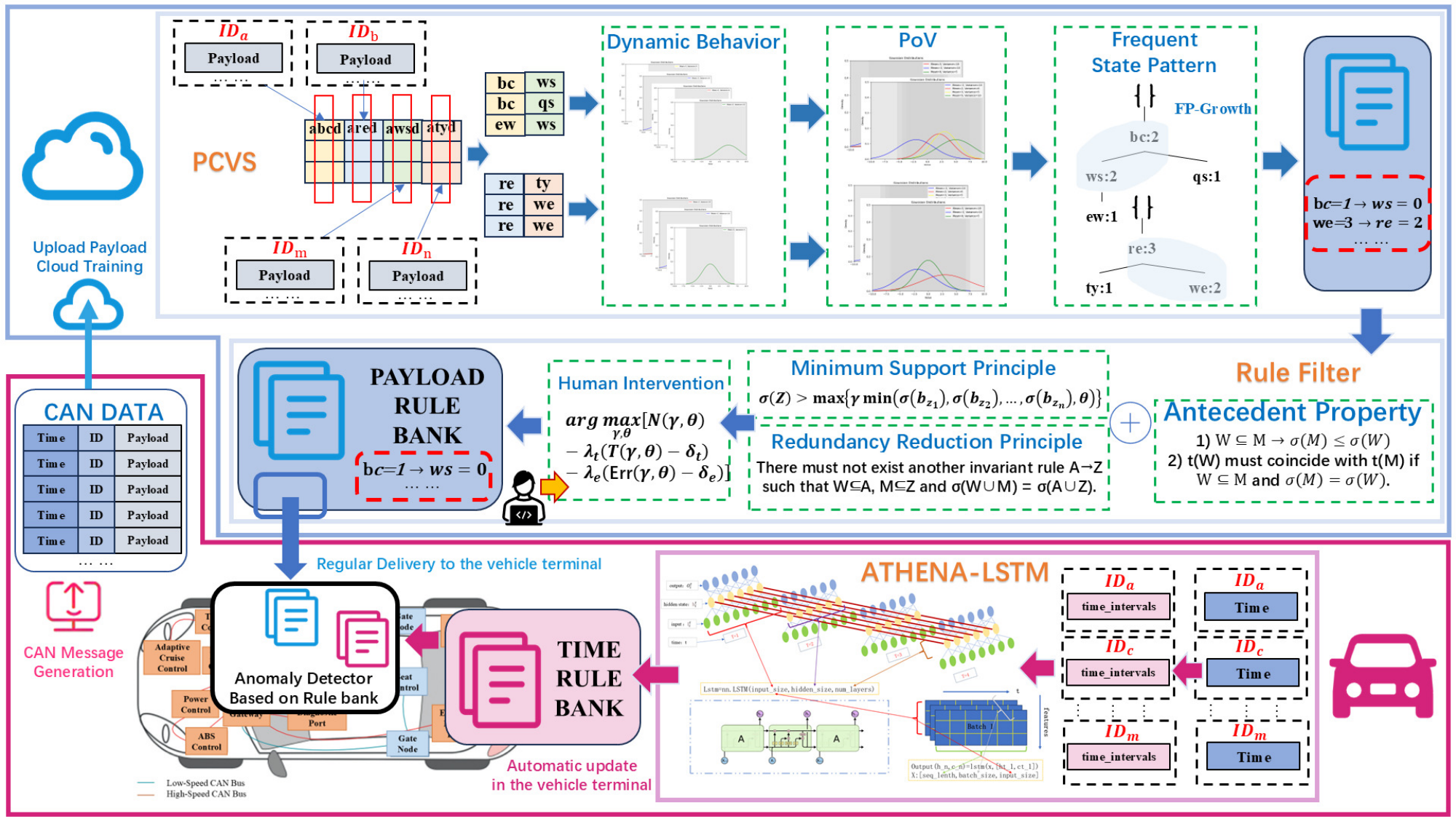"}
    \caption{ATHENA Framework.}
    \label{fig_3}
\end{figure*}

ATHENA consists of the following two components, as illustrated in Figure \ref{fig_3}.
\begin{itemize}
\item \textbf{Vehicle-Cloud Processing and Verification Module (PCVS)} (\ref{sec4.2}) integrates the periodic clustering characteristics of CAN messages and the self-supervised detection mechanism based on the physical characteristics carried by CAN messages of the vehicle system. The payload is uploaded to the cloud-based PCVS for training, considering vehicle-cloud computational power differences. By focusing on the distribution of CAN message data with different IDs within the same periodic cluster, PCVS forms a stable payload rule bank that reflects the physical laws of the vehicle system under specific time and space conditions. This rule bank ensures that the information updates transmitted by the ECU are directly constrained by the overall dynamic behavior of the physical system of the vehicle, thus maintaining the accuracy and responsiveness of the vehicle control.

\item \textbf{ATHENA-LSTM} (\ref{sec4.3}) is an innovative method based on LSTM and IVNs. It learns the "normal sending pattern" of each ID CAN message to generate a time rule bank composed of temporal rules, and then performs intrusion detection through rule-based method. Instead of using LSTM for detection only after training LSTM. The method first normalizes the sending intervals of normal CAN messages to amplify small gaps caused by short periodicity. These processed timing features are then fed into an LSTM for training. The adam optimizer was used as the loss function in the training process. After the training was completed, the Mean Square Error (MSE) was used to adjust the prediction results to create a dynamic time rule bank reflecting the normal sending patter, which can adapt to varying data distributions, improving the detection of Timing Transparent attacks in the ATHENA framework, while leveraging the physical characteristics of the vehicle system.
\end{itemize}

\subsection{PCVS}\label{sec4.2}

First of all, ATHENA abandoned the concept that most of the existing research takes the continuous arrival message set as the exploration basis, but takes all the messages with the same identifier as a probe unit (such as all the CAN messages with the same ID), and then divides the timestamp and the payload of this probe unit. Then, combined with the vehicle-cloud integrated environment and the difference of computing power between the vehicle and the cloud, the payload was uploaded to the cloud PCVS for model training. 
%突出不同点
We refer to the study of Cheng Feng et al.\cite{A17}, using physical characteristics to guide rule generation in the field of IVN intrusion detection, which is different from its original industrial control application scenario with completely different physical operating mechanisms. 
PCVS integrates the periodic clustering characteristics of CAN messages with a general self-supervised detection mechanism based on the physical characteristics of CAN messages carried by vehicle systems. It relies on the pattern change characteristics of byte information reflecting physical characteristics of CAN communication data to identify attacks, thus avoiding complex protocol reverse engineering and message semantic parsing process. It provides a general intrusion detection framework for different communication matrix field definitions of different vehicle manufacturers. In short, from the perspective of data distribution, taking the CAN communication matrix field definition as a reference, the local bit spatio-temporal change results of CAN message data with different IDs in the same periodic cluster are diverted into the distribution domain that reflects the operation law of the vehicle system from different angles in a multidimensional direction. Then a kind of data sample set with stable internal distribution is formed, which can reflect the physical law of the system under specific time and space. Later, in the vehicle system, the information update transmitted by the ECU at each time step is usually directly affected by \textit{\uline{the overall dynamic \textbf{B}ehavior \textbf{O}f the current \textbf{V}ehicle physical system (BoV)}}. This mechanism reflects the first-principled physical knowledge of the vehicle system. For example, the ECU controlling the wheel speed may be triggered by multiple BoVs such as acceleration and braking, which change the wheel speed control requirements in real time and guide the ECU to adjust the output information to adapt to BoV. This physical-event-driven information update mechanism ensures the accuracy and responsiveness of the vehicle control system.

\begin{figure}[!ht]
    \centering
    \includegraphics[height=9cm]{"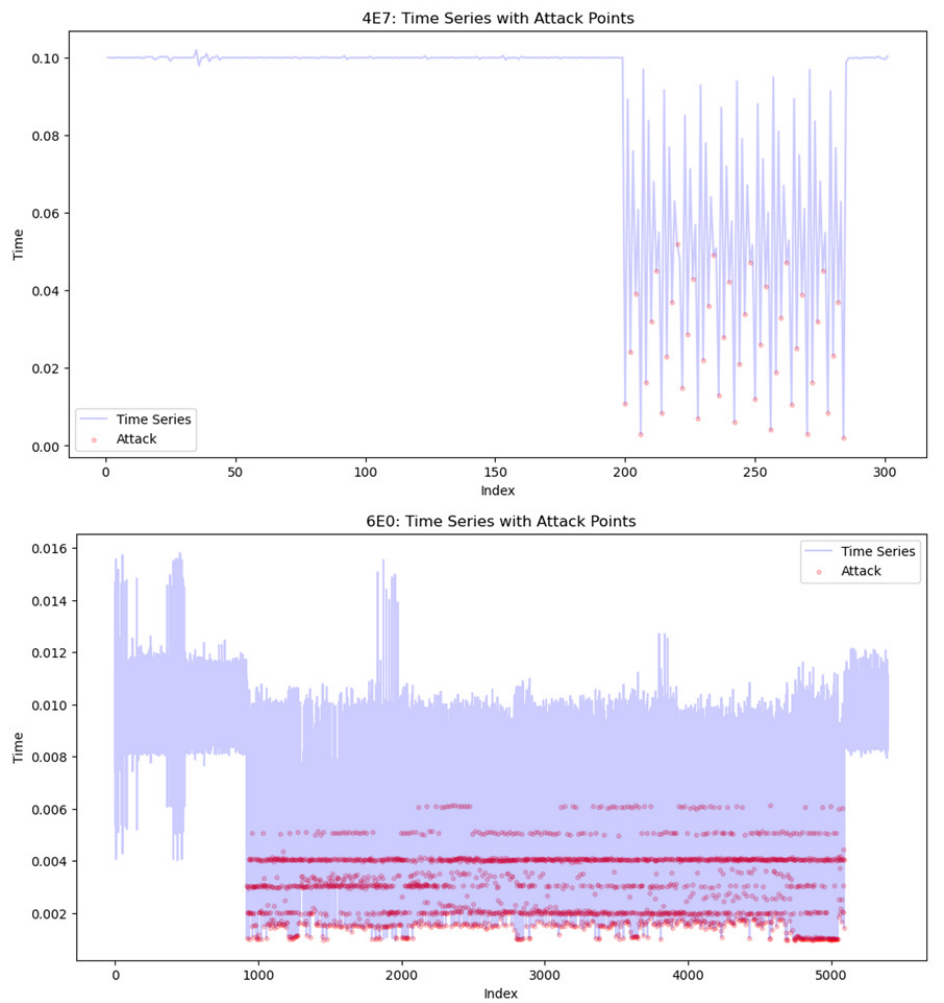"}
    \caption{CAN Time Series.}
    \label{fig_4}
\end{figure}

Before detailing the PCVS module, we present the periodic clustering characteristics of CAN messages that we found. CAN messages have different sending periodicity according to different IDs, and this sending periodicity will be affected by the BoV of the current vehicle system. For instance, the periodicity can change when the engine is in ignition or shutdown. In the short term, the periodicity is fixed, while in the long term, it varies within a range. As shown in Figure \ref{fig_4}, these are two sets of time series of CAN messages that have been attacked by the Timing Transparent type attack, where the abscissa represents the number of messages and the ordinate represents the timing information. The red dot represents the point where the attack occurs. It can be seen from the figure that the periodicity for ID 4E7 remains constant at 0.10s in the short term, while ID 6E0 shows periodicity changes within a range over a longer period. By grouping IDs based on periodicity (periodic clusters), we can better understand the coordination between ECUs and reduce the complexity of state pattern mining.

Combining the set of data samples reflecting the numerical distribution of CAN messages for a given ID, the characteristics of periodic clusters, and the above first-principled physical knowledge of the vehicle system, \textit{\uline{the \textbf{P}hysical state information set \textbf{O}f the \textbf{V}ehicle system (PoV)}} $\{\Delta \mathbf{x}^t \sim N_1, \dots, \Delta \mathbf{x}^t \sim N_k\}$ for the given ID can be generated. After generating the above PoVs for each ID in the same periodic cluster, the FP-Growth algorithm\cite{A66} is used to mine the frequent state patterns between the PoVs of different IDs, and an initial rule bank satisfying a certain support condition is generated. Then, the rule bank is streamlined and optimized by the Minimum Support Principle, the Redundancy Reduction Principle and the Human Intervention Principle. Finally, a set of characteristic payload rule bank is formed.

The BoV is composed of the dynamic behavior of the individual hardware systems together, and the PoVs corresponding to different IDs of CAN messages is the numerical embodiment of "dynamic behavior of each hardware system". Each distribution in the set represents a BoV, and the numerical changes across different BoVs vary. First, let $\text{Data}^{\{1:t\}} = \{d^1, d^2, \dots, d^T\}$ is a time series data log of CAN messages (no abnormal signal), $B = \{b_1, b_2, \dots, b_k\}$ is a set of $k$ dynamic behaviors, and each $d^t \in \text{Data}^{\{1:t\}}$ satisfies a subset of dynamic behaviors in $B$. This subset CAN be represented by $B^t$, $B^t \in B$, which is used to represent the set of states of the vehicle system at this time in a specific time interval.

The information update transmitted by the ECU at each time step is usually directly affected by the BoV. It is recorded that at time $t$, the data reading transmitted by the ECU is $x^t$. Then $\Delta x^t = x^{t+1} - x^t$ represents the update of the data reading transmitted by the ECU from the time interval $t$ to $t+1$. Define the BoV $k \in (1, 2, \dots, K)$, which means that the physical system of the vehicle is in the $k'th$ state of the $K$ kinds of BoVs. For example, state $k$ can be acceleration or braking at this time. The amount of information update at any single time instant is defined as $\Delta x^t$, which is expressed as follows:
\begin{equation}
\Delta x^t = \mu_k + \epsilon_k
\label{eq:1}
\end{equation}
Where $\mu_k \in \mathbb{R}$ represents the expected update of the corresponding ECU reading when the physical system of the vehicle is in state $k$, and the noise $\epsilon_k \sim \mathcal{N}(0, \sigma_k^2)$ is a random process normally distributed with zero mean. All updates of CAN readings in different states of the same ID passed by the ECU are denoted as $\Delta x^{\{1:T-1\}} \equiv \{\Delta x^1, \Delta x^2, \dots, \Delta x^{T-1}\}$ corresponds to a kind of BoV, which is generated by $K$ different distributions with unknown parameters.

Once the distribution of vehicular network communication data for the same CAN ID across different time and spatial domains is inferred, the following PoV can be generated for the system physical characteristics of the CAN ID in different time and space domains: $\{\Delta x^t \sim N_1, \dots, \Delta x^t \sim N_k\}$, where the physical state information $\Delta x^t \sim N_k$ of the vehicle system indicates that the update of the data at time step $t$ is generated from the $k'th$ distribution, and the BoV is in state $k$ at this time.

To generate the above PoV for each CAN ID, it is first necessary to infer $K$ distributions that generate $\Delta x^{\{1:T-1\}}$. We use the Expectation-Maximization (EM) algorithm\cite{A67} to fit the Gaussian Mixture Modules (GMMs)\cite{A68} with $K$ distributions to generate $\Delta x^{\{1:T-1\}}$ to infer their distributions\cite{A17,A33}. Specifically, a GMM with $K$ components is defined by three parameter vectors: average $\mu = \{\mu_1, \dots, \mu_k\}$ and standard deviation $\sigma = \{\sigma_1, \dots, \sigma_k\}$ (which determines the CAN data to be updated in $K$ different BoVs), and the mixture weight vector $\pi = \{\pi_1, \dots, \pi_k\}$ of the $K$ components (which represents the prior probability that a CAN data update belongs to each BoV). A set of fitting candidate GMMs with different distribution numbers is calculated, and the PoV is generated by the minimum Bayesian Information Criterion (BIC) score\cite{A17,A34}. This is considering that the BIC has a balance between the likelihood function of the model and the complexity of the model, which is more suitable for model selection in the context of the limited set of models in this topic. The corresponding BIC score of a candidate GMM is defined by the following formula\cite{A34}:
\begin{equation}
BIC(\mathcal{M}_K) = -2 \log p(\Delta X^{\{1:T-1\}} \mid \mathcal{M}_K) + \kappa \log(n)
\label{eq:2}
\end{equation}

Therefore, based on the above selection of the physical state information $\Delta x^t \sim N_k$ of the vehicle system is satisfied at time $t$, if at time $t$ and only if:
\begin{equation}
r_k^t = \max(r_1^t, \dots, r_k^t)
\label{eq:3}
\end{equation}

$r_k^t$ denotes the membership probability of the $k'th$ distribution to the ECU passing the information update $\Delta x^t$ at time $t$, which is calculated by the following formula:
\begin{equation}
r_k^t = \frac{\pi_k N(\Delta x^t \mid \mu_k, \sigma_k)}
{\sum_{j=1}^K \pi_j N(\Delta x^t \mid \mu_j, \sigma_j)}
\label{eq:4}
\end{equation}

After we generate the PoVs for all IDs in the same periodic cluster, we need to deeply mine the association between different IDs in a certain time interval, which reflects a cooperative operation rule of the vehicle system. Here, we use the FP-Growth algorithm to mine the combination of frequently occurring dynamic behaviors of different IDs. The so-called combination of frequently occurring dynamic behaviors of different IDs is the combination of physical state information of the vehicle system with different IDs according to the pre-determined support threshold, which is called the state pattern. To illustrate the concept of support, let $B = \{b_1, b_2, \dots, b_k\}$ is the set of $k$ dynamic behaviors. Each $d^t \in \text{Data}^{\{1:t\}}$ satisfies a subset of the dynamic behavior in $B$, which can be denoted by $B^t$,  $B^t \in B$, which is used to represent the set of states that CAN message data is in at this time in the vehicle system in a specific time interval. Each $d^t \in \text{Data}^{\{1:t\}}$ satisfies a subset of the dynamic behaviors in $B$, denoted as $B^t$, $B^t \in B$, which is used to represent the set of the set of states that CAN message data occupies within the vehicle system during a specified time interval. Let $\sigma(X)$ denote the support of a state pattern $X$, which is the fraction of time steps for which the state pattern $X$ is contained by $B^t$, and is calculated as follows\cite{A17}.
\begin{equation}
\sigma(X) = \frac{\sum_{t=1}^T 1(X \subseteq I^t)}{T}
\label{eq:6}
\end{equation}

The definition of support mainly declares the timeliness of the state pattern, and adds the consideration of time characteristics to the value judgment of the state pattern. To illustrate the process of discovering frequent state patterns, after we generate the corresponding PoV for different IDs $\{\Delta x_A^1 \sim N_1, \dots, \Delta x_A^t \sim N_k\}$, we want to explore the combination of dynamic behaviors that may have strong ties between different IDs, that is, state patterns.

After exploring the frequent state pattern, what we need to consider is the generated rules, that is to say, for a set of meaningful frequent state pattern $\{\Delta x_A^1 \sim N_1, \Delta x_B^1 \sim N_3\}$ is $\Delta x_A^1 \sim N_1 \to \Delta x_B^1 \sim N_3$ or $\Delta x_B^1 \sim N_3 \to \Delta x_A^1 \sim N_1$.

We define invariant rules as follows:
\begin{equation}
X \Rightarrow Y \iff X \cap Y = \emptyset \land \sigma(X \cup Y) = \sigma(X)
\label{eq:7}
\end{equation}

Where $X$ is called the antecedent and $Y$ is called the consequent of the rule. This rule means that when a set of dynamic behaviors $X$ appears as antecedents in a set of state patterns, another set of state patterns $Y$ appears accordingly, and $X$ and $Y$ are mutually exclusive.

For example, if the CAN ID representing the angle of the accelerator pedal is $A$ and its payload is $DA$, and the CAN ID of the brake pedal angle is $B$ and its payload is $DB$, then the invariant rule: $\{A, DA \downarrow\} \Rightarrow \{B, DB \to Max\}$, means that if  the dynamic behavior $\{A, DA \downarrow\}$ is satisfied at any given time step $t$, the dynamic behavior $\{B, DB \to Max\}$  must also be satisfied at $t$. This example reflects that in the acceleration process of the actual operation of the vehicle system (A kind of BoV), when the accelerator pedal angle decreases, the brake pedal angle is at its maximum value. In this derivation process, both antecedent and consequent are true values, so the derivation is also true, so this derivation can be used as a rule that the system needs to follow under the overall dynamic behavior of "normal driving". Such rules reflect a deep entanglement relationship of the physical state of the vehicle system, which is not artificially prescribed, but a symbolic representation of the operation law of the vehicle system itself under an overall dynamic behavior. This also confirms the first-principled physics knowledge of the vehicle system we mentioned, which is specifically expressed in the form of rules for intrusion detection in IVNs.

Following the previous analyses, we now focus on the criteria that define the meaningfulness of a set of rules. For a set of rules to attain meaningfulness, they must not only demonstrate statistical significance but also exhibit relevance to the task at hand. Additionally, the derived rules should possess mathematical rigor and practical utility. The meaningful rules in this study are mainly constrained by three principles, which are: Minimum Support Principle, the Redundancy Reduction Principle and the Human Intervention Principle. Before doing so, we want to introduce some properties that can be rigorously proved using discrete mathematics:

\noindent\textbf{Property 1.} $W \subseteq M \to \sigma(M) \leq \sigma(W)$

\noindent\textbf{Property 2.} Let $t(W)$ denote the set of time steps at $W \in B^t$, then t(W) must coincide with $t(M)$ if $W \subseteq M$ and $\sigma(M) = \sigma(W)$.

Specifically, the first precondition illustrates the anti-monotonicity property possessed by rule mining in this study, which states that if the state pattern $W$ is a subset of $M$, then $\sigma(M)$ does not exceed $\sigma(M)$. So the second precondition must also hold.

Under the constraints of the above preconditions, we introduce three constraint principles for generating rules.

If the invariant rule $W \to M$ is meaningful, then Minimum Support Principle is formally defined as follows\cite{A17}:
\begin{equation}
\sigma(Z) > \max\{\gamma \min(\sigma(b_{z_1}), \sigma(b_{z_2}), \dots, \sigma(b_{z_n})), \theta\}
\label{eq:8}
\end{equation}

Where $Z = W \cup M$, $\{b_{z_1}, b_{z_2}, \dots, b_{z_n}\}$ denotes all terms in $Z$.

And according to the anti-monotonicity property, it can be seen that the support of an invariant rule is constrained by the support of the least frequent item in it:
\begin{equation}
\sigma(Z) \leq \min\left(\sigma(b_{z_1}), \sigma(b_{z_2}), \dots, \sigma(b_{z_n})\right)
\label{eq:9}
\end{equation}
The first principle ensures that the invariant rules achieve the required statistical significance, which allows the derived rules to be satisfied not just by chance in the datalog, since incidentally satisfied rules cause a lot of false positives when used for anomaly detection. Then, since the support for different state patterns CAN vary greatly in CAN data log, for example, in the running vehicle system data log, the brake pedal angle only varies in 10\% of the time steps. It is therefore unfair to set a unique minimum support threshold for all invariant rules. We require that the support of an invariant rule be greater than the product of $\gamma$ with its particular upper bound, where $\gamma \in (0, 1)$. Moreover, $\theta \in (0, \gamma)$ is another threshold that controls the minimum fraction of samples in the data log that a meaningful invariant rule needs to capture, which guarantees that those rare items are excluded from any meaningful invariant rule.

Next, the Redundancy Reduction Principle is stated, which is formally defined as follows.

For an invariant rule $W \to M$ to be meaningful, then there must not exist another invariant rule $A \to Z$ such that $W \subseteq A$, $M \subseteq Z$ and $\sigma(W \cup M) = \sigma(A \cup Z)$.
The second principle reduces the derived invariant rules because a large number of redundant invariant rules does not improve the performance of the anomaly detection model, but it significantly increases the time cost of the anomaly detection process. Specifically, A rule that does not satisfy the above principle is redundant, since if there exists $W \cup M \subseteq A \cup Z$ with $\sigma(W \cup M) = \sigma(A \cup Z)$, then by Property 2, $t(W \cup M)$ must overlap $\to t(A \cup Z)$. This means that if $W \to M$ is violated at any given time step, then $A \to Z$ must also be violated. Thus, the rule $W \to M$ will have no additional contribution beyond the rule $A \to Z$ during anomaly detection.

After that, we integrate the Human Intervention Principle into the rule mining, which greatly improves the generalization ability of the model. This principle makes the mined invariant rules not only obey the original rules guided by the first-principled physical knowledge of the vehicle system, but also be affected by the subjective needs of users, which makes the rule generation process more flexible and variable. In the invariant rule mining phase, the number of meaningful invariant rules generated according to the Minimum Support Principle and the Redundancy Reduction Principle is affected by the values of two important parameters $\gamma$ and $\theta$ in the following formula, which define the rule level and the global minimum support threshold of the generated rules, respectively. Specifically, smaller values of $\gamma$ and $\theta$ make the generated invariant rules more meaningful, and therefore the potential chance of using these rules to reveal anomalies increases. However, the statistical significance of the derived rules will also decrease, which will potentially lead to more false positives when using the rules for anomaly detection. In addition, the processing time cost of checking the invariant rule at each data point increases.

The formal definition of the Human Intervention Principle is now given as follows\cite{A17}:
\begin{equation}
\arg\max_{\gamma, \theta} \left[N(\gamma, \theta) - \lambda_t \left(T(\gamma, \theta) - \delta_t\right) - \lambda_e \left(\text{Err}(\gamma, \theta) - \delta_e\right)\right]
\label{eq:10}
\end{equation}

In this formulation, $T(\gamma, \theta)$ is the time cost of anomaly detection with invariant rules on each data point, $N(\gamma, \theta)$ is the number of meaningful invariant rules generated by $\gamma$ and $\theta$, and $\delta_t$ and $\delta_e$ are user-defined thresholds for acceptable time cost and acceptable validation error of processing each data point. $\lambda_t$ and $\lambda_e$ are the weights of the time constraint penalty term and the error constraint penalty term, respectively, with $\lambda_t > 0$ and $\lambda_e > 0$. What this definition expresses is maximizing the number of meaningful invariant rules generated at an acceptable time cost for processing each data point and validation error. At this point, the generation of the payload rule bank which can reflect the operation rule of the vehicle system under the guidance of the first-principled physical knowledge of the vehicle system is completed.

Finally, the constructed payload rule bank should be delivered to the vehicle, and the intrusion detection of IVNs should be carried out according to the payload rule bank at the vehicle. When detecting, in a time step, if the dynamic behavior corresponding to the information update $\Delta x^t$ transmitted by an ECU is generated under the BoV $k$, then it must satisfy the distribution under BoV $k$, that is, $\Delta x^{t'} \sim \mathcal{N}^{k+1}$. Moreover, let $\Delta x^{t'}$ be the information update of one ECU in the detection phase, we define $\Delta x^{t'} \sim \mathcal{N}^{k+1}$, but if the following situation arises:
\begin{equation}
r^{t'}_k < \min(r^1_k, \ldots, r^{T-1}_k), \quad \forall k \in (1, \ldots, K)
\label{eq:11}
\end{equation}

This means that $\Delta x^{t'}$ is an anomalous ECU update that does not produce such an information update under any BoV and does not produce such a distribution of message values in a time interval, so it is treated as an anomaly. It can be seen that the generation of rules is very complex, but the detection of anomalies is very fast, which is the advantage of Rule/Specification-Based detection. This method also makes full use of the abundant computing resources of the cloud and the relatively lack of computing power of the vehicle terminal.

\subsection{ATHENA-LSTM}\label{sec4.3}

As mentioned in section \S \ref{sec4.2}, the transmission of CAN data usually has a significant periodic rule, which can be used as an important basis for intrusion detection and can efficiently detect those Transparent Timing attacks. However, in the actual operation of the vehicle, the sending periodicity of the CAN message is not invariable, which has been introduced in section \S \ref{sec4.2}. Therefore, our research uses the LSTM model, which is excellent in time series prediction tasks, to solve this problem. In addition, the lightweight structure of LSTM can also be well trained autonomously at the vehicle terminal. ATHENA-LSTM is a new method based on LSTM and IVNs. It learns the "normal sending pattern" of each ID CAN message to generate a time rule bank composed of temporal rules, and performs intrusion detection through rule-based method. Instead of using LSTM for detection only after training LSTM\cite{A60,A61,A62}.

\begin{figure}[!ht]
    \centering
    \includegraphics[height=5cm]{"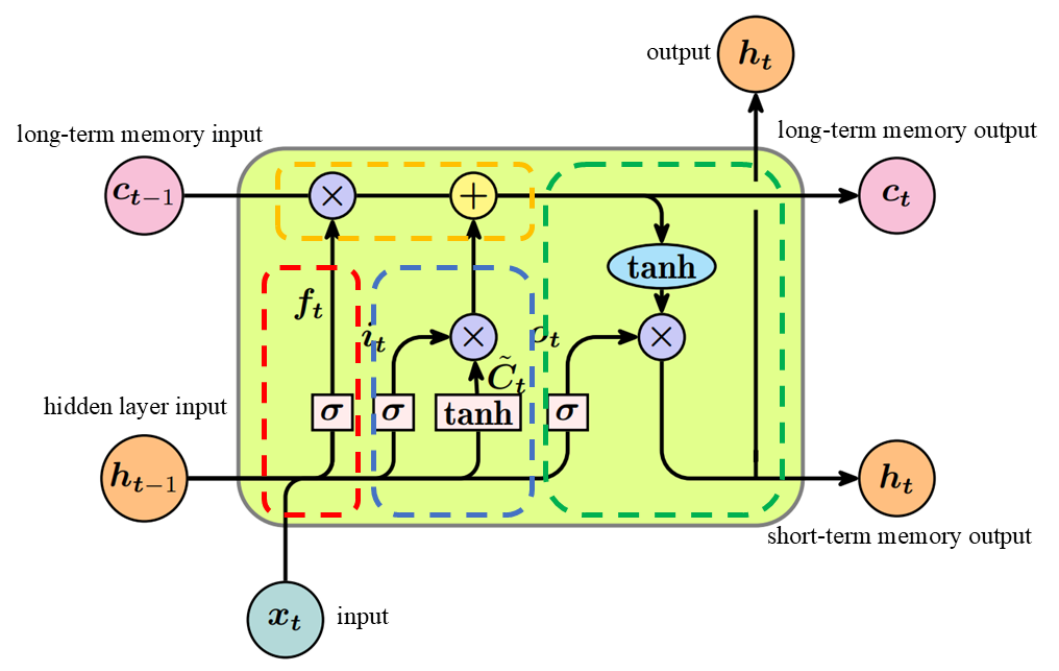"}
    \caption{LSTM Unit.}
    \label{fig_5}
\end{figure}

As we all know, Recurrent Neural Network (RNN) is a kind of neural network with memory ability. In RNN, the neurons can not only learn the features of the input data at the current time, but also receive the features of the previous input data from the neurons at the previous time to form a long-term "memory". LSTM is a variant of RNN. While effectively solving the problem that the original RNN is prone to gradient explosion or gradient disappearance during training, it introduces the Input Gate $i_t$ (blue box), the Forget Gate $f_t$ (red box), the Output Gate $f_t$ (green box) and a Cell State (yellow box). These mechanisms allow LSTM to better handle long-term dependencies in time series as shown in Figure \ref{fig_5}.

\begin{figure}[!ht]
    \centering
    \includegraphics[height=5cm]{"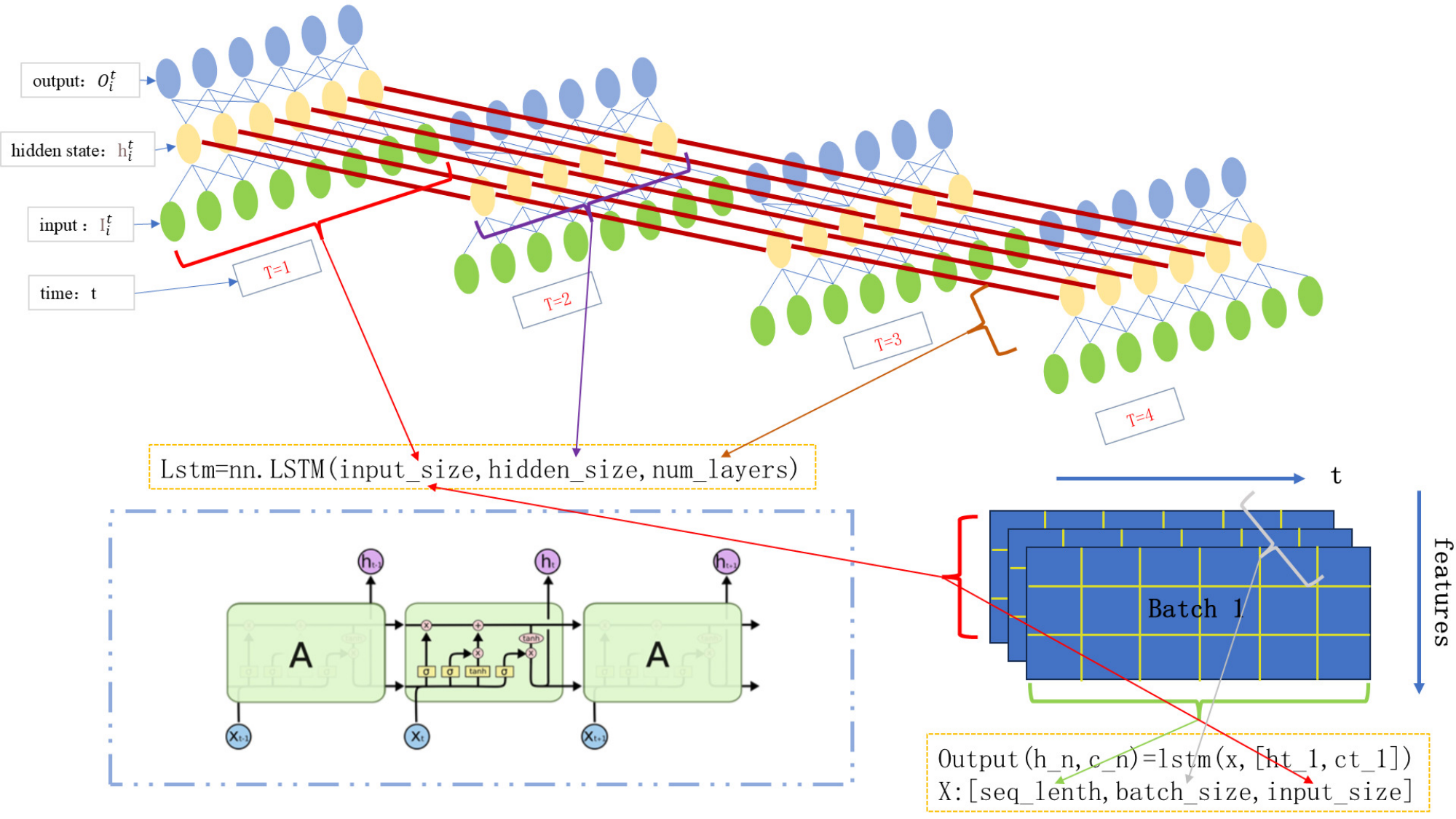"}
    \caption{LSTM Model.}
    \label{fig_6}
\end{figure}

Let $W$ be the weight matrix of each gate, $b$ be the calculation bias of each gate, $\tilde{C}_t$ be the cell state of current memory cell $C_t$, then the specific calculation process of LSTM is shown in equations (12) to (17).
\begin{equation}
f_t = \sigma(W_f * [h_{t-1}, x_t] + b_f)
\label{eq:12}
\end{equation}
\begin{equation}
i_t = \sigma(W_i * [h_{t-1}, x_t] + b_i)
\label{eq:13}
\end{equation}
\begin{equation}
o_t = \sigma(W_o * [h_{t-1}, x_t] + b_o)
\label{eq:14}
\end{equation}
\begin{equation}
\tilde{C}_t = \tanh(W_C * [h_{t-1}, x_t] + b_C)
\label{eq:15}
\end{equation}
\begin{equation}
C_t = f_t * C_{t-1} + i_t * \tilde{C}_t
\label{eq:16}
\end{equation}
\begin{equation}
h_t = o_t * \tanh(C_t)
\label{eq:17}
\end{equation}
The overall structure of LSTM adopted in this study is shown in Figure \ref{fig_6}. The timing characteristics of the physical state information of the vehicle system contained in CAN messages have been deeply analyzed in section \S \ref{sec4.2}. In this part, we propose an innovative method to process the timing characteristics of CAN messages. Instead of directly taking all timing features of the same ID as the input of LSTM to predict the timing of subsequent CAN messages, our goal is to learn a set of correct sending time intervals for each ID through LSTM, which is called "normal sending pattern".

Specifically, for a set of timing characteristics of normal CAN messages, we first calculate the sending interval of all messages and normalize it to scale the sending interval to the 0-1 interval. Since many CAN messages are sent with extremely short periodicity, resulting in small gaps between sending intervals, this step aims to amplify these gaps to improve the effect of model training. Subsequently, the processed temporal features are fed into the LSTM for training. After training, by subtracting two times of the standard deviation from the prediction results (compared with the absolute error, the standard deviation can better adapt to the data with large changes in distribution), the normal sending pattern is generated, which is the time rule bank. Finally, the time rule bank is used for rule-based intrusion detection, instead of simply training LSTM and using the trained LSTM model for detection. At the same time, the relatively time-consuming LSTM training process and the rule-based real-time classification task are decoupled.

The time rule bank based on ATHENA-LSTM can dynamically adapt to the periodic fluctuations of CAN messages instead of relying on a fixed detection threshold. Our data processing method innovates to capture the timing characteristics of CAN messages, effectively enhancing the efficiency of ATHENA framework in detecting Timing Transparent attacks, while highlighting the uniqueness of this research guided by the physical characteristics of vehicle systems.

\section{Experiment}\label{section5}
In this section, we implement the ATHENA framework in environment \S \ref{sec5.1} and present the evaluation metrics in \S \ref{sec5.2}. Typical benchmark methods for horizontal comparison are given in \S \ref{sec5.3} and details of the ROAD dataset are given in \S \ref{sec5.4}. Finally, a large number of experiments on running performance and detection performance are carried out in \S \ref{sec5.5}, as well as the result analysis. The impact of hyperparameters on model performance is given in \S \ref{sec5.5}. Finally, the results and analysis of ablation experiments are presented in \S \ref{sec5.6}.

\subsection{Environment}\label{sec5.1}
With the rise of Internet technology and 4G/5G networks, vehicles are increasingly connected to the Internet, driving vehicle-cloud integration.  Major automakers and service providers are focusing on building highly abstract vehicle-cloud architectures, promoting seamless collaboration between the vehicle terminal and cloud services.  This "vehicle-cloud integrated architecture" allows flexible invocation of cloud services by vehicles and efficient utilization of vehicle capabilities by the cloud, creating a new ecosystem where "cloud empowers cars, and cars serve people."

ATHENA framework integrates the vehicle-cloud integrated architecture, which combines the computation-intensive non-real-time rule mining on the cloud with the real-time rule-based rapid detection on the vehicle, fully realizes the efficient scheduling and coordination of the vehicle and cloud capabilities, and successfully integrates the functions of the vehicle and the cloud organically.

\begin{table}[ht]
\centering
\label{device_config}
\small % 缩小字体
\renewcommand{\arraystretch}{1.5} % 调整行间距
\setlength{\tabcolsep}{3pt} % 缩小列间距
\rowcolors{2}{white}{green!10} % 设置斑马纹
\caption{Device Configuration (Vehicle-Grade Device and Cloud)}
% 使用 p{宽度} 来控制列宽
\begin{tabular}{|>{\centering\arraybackslash}m{2cm}|>{\centering\arraybackslash}m{3cm}|>{\centering\arraybackslash}m{3cm}|} % 设置居中
\hline
\rowcolor[HTML]{C1E0FF} 
\label{ATHENA_Device}
\textbf{Environment} & \textbf{Vehicle Terminal} & \textbf{Cloud Terminal} \\ \hline
Type of System & Jetson Orin Nano & ROG Strix G814JV\_G814JV \\ \hline
OS & Ubuntu 20.04 & Microsoft Windows 11 \\ \hline
CPU & ARMv8 

Processor rev 1 & 13th Gen Intel(R) Core(TM)i9-13980HX,2200Mhz\\ \hline
GPU & 512-core NVIDIA Ampere architecture 

GPU & Geforce RTX 4060 \newline 32 GB \\ \hline
RAM & 8.0 GB & 32.0 GB \\ \hline
\end{tabular}   
\end{table}
Our cloud devices and car devices are shown in Table \ref{ATHENA_Device}. It is worth noting that the vehicle terminal simulation device is TW-T206 from TWOWIN Technology, China. The CPU and GPU share 8 GB RAM, and the computing power is only 40 TOPS, which simulates the limited computing power of the vehicle terminal.
\begin{figure}[!ht]
    \centering
    \includegraphics[height=3cm]{"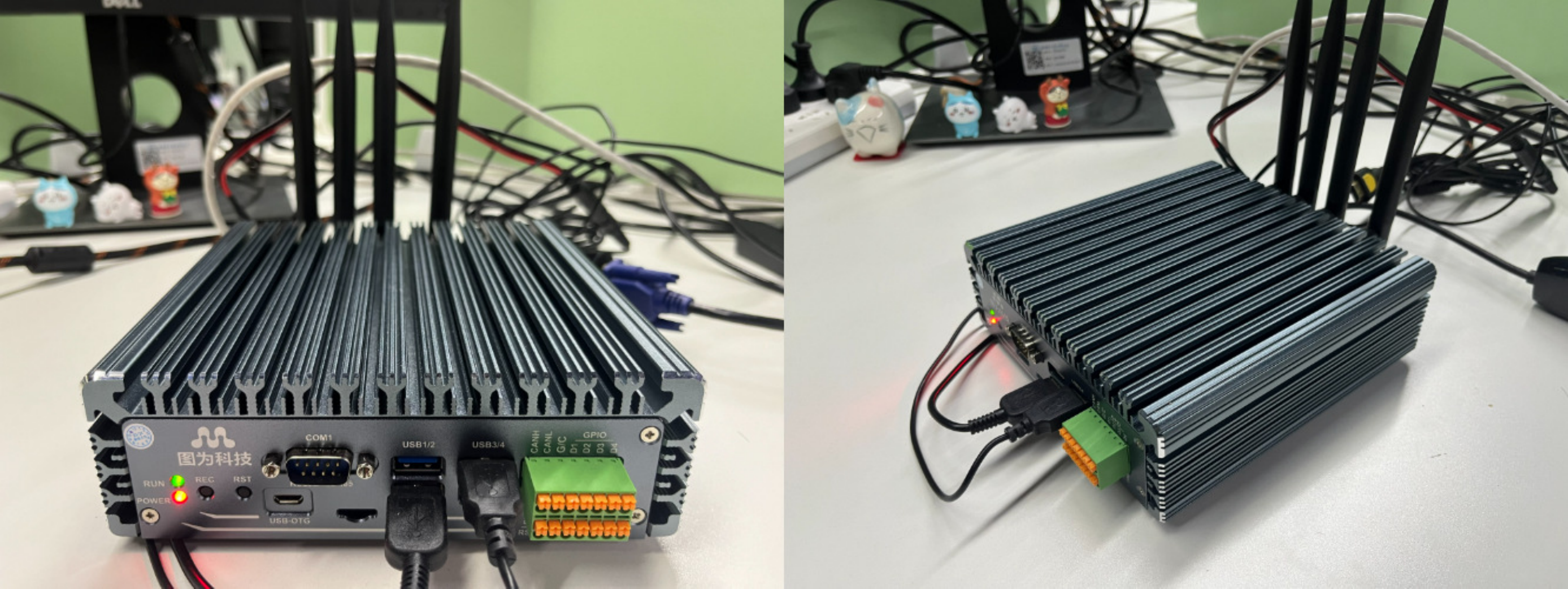"}
    \caption{On-board Computing Unit.}
    \label{fig_7}
\end{figure}
\subsection{Evaluation metrics}\label{sec5.2}
In order to evaluate the effectiveness of different methods, the classical evaluation method is used in this research. Accuracy, Precision and Recall can be calculated by using the number of true negative (TN), true positive (TP), false negative (TN) and false positive (FP), defined as follow:
\begin{equation}
\text{Accuracy} = \frac{TP + TN}{TP + TN + FP + FN}
\tag{18}
\end{equation}
\[
\text{Precision} = \frac{TP}{TP + FP}
\tag{19}
\]
\[
\text{Recall} = \frac{TP}{TP + FN}
\tag{20}
\]
Moreover, F1-score is more useful than accuracy in the case of imbalanced class distribution. It is the harmonic mean of precision and recall and is calculated as follows:
\[
F1\text{-score} = \frac{2 \times \text{Precision} \times \text{Recall}}{\text{Precision} + \text{Recall}}
\tag{20}
\]

We also use the ROC curve and AUC to evaluate model performance. The ROC curve shows the trade-off between true and false positive rates, while AUC reflects the model’s discriminatory power. These metrics ensure a comprehensive assessment of model stability.

\subsection{Compared Methods}\label{sec5.3}

To verify ATHENA’s performance and conduct a comprehensive comparison, we selected several classical methods as benchmarks, including lightweight networks (e.g., MobileNetV3 and EfficientNet), intrusion detection models for IVNs (e.g., CANet, CAN-RF, CAN-MLP, CAN-LSTM), and graph-based IDS (e.g., Graph-based IDS and G-IDCS TH classifier). These models represent a range of approaches, from CNN and time series modeling to random forests and graph neural networks, offering a multi-dimensional basis for performance comparison.
\begin{itemize}
    \item \textit{MobileNetV3}\cite{A46}: MobileNetV3 is a lightweight deep neural network optimized by NAS and SE modules, which combines low computational cost and excellent performance, making it ideal for edge computing and real-time applications.
    \item \textit{CANet}\cite{A45}: This is the first IVN multi-class intrusion detection method based on deep learning, which examines a single CAN message within each detection window through a newly designed network structure. The network architecture is specifically designed for the signal space of CAN message, where each signal within the payload region of CAN data is learned by a single RNN unit.
    \item \textit{CAN-RF}\cite{A47}: It is an intrusion detection method based on the random forest algorithm and focuses on CAN message security in IVNs. By extracting the features of CAN messages, CAN-RF uses random forest model for classification and anomaly detection, which CAN effectively identify potential network attacks. CAN-RF is often used as a lightweight and low-latency intrusion detection solution, which is widely used in the field of Internet of vehicles security.
    \item \textit{MultiLayer Perceptron(CAN-MLP)}\cite{A47}: It is an intrusion detection method based on Multilayer Perceptron (MLP), which has the characteristics of simple structure and high computational efficiency. It uses a feedforward neural network structure to perform multiple classification tasks for each individual CAN message.
    \item \textit{EfficientNet}\cite{A48}: This is an efficient CNN. By introducing a compound scaling strategy and adjusting the depth, width and resolution of the network at the same time, it improves the accuracy while significantly reducing the computational cost, and can perform multi-classification intrusion detection tasks in the intrusion detection environment with limited computing power.
    \item \textit{G-IDCS TH classifier}\cite{A49}: It is a key component of a graph-based IDS that focuses on the classification and anomaly detection of network traffic using a Threshold Classifier (TH). It analyzes communication patterns through graph structure and combines threshold rules to quickly and efficiently identify potential intrusions.
\end{itemize}

\subsection{Datasets Details}\label{sec5.4}
Seven publicly available CAN datasets with labeled attacks were analyzed to explain our choice of the ROAD dataset for experiments. The CAN Intrusion (OTIDS)\cite{A41} dataset contains fuzzing attacks but lacks labeled attack messages, and its interval documentation is unclear. The Survival Analysis Dataset\cite{A42} has simple attacks on multiple vehicles, easily detectable with our LSTM module. The Car Hacking Dataset\cite{A42,A44}, though real, has long capture periods and message gaps, making it inconsistent with actual vehicle behavior. The SynCAN Dataset\cite{A45} is simulated and, while useful, doesn't perfectly replicate real data. The Automotive CAN Bus Intrusion Dataset v2 includes real and simulated data with diagnostic protocol and simulated suspension attacks. The Can Log Infector framework simulates impersonation attacks, but post-processing may not affect vehicle functionality. The ROAD dataset, proposed by Oak Ridge National Laboratory, is the first to include real, advanced attacks, offering a more realistic and diverse attack set, making it our choice for experiments. In addition to the usual attack types, the ROAD dataset includes the following four masquerade attacks: 

\begin{itemize}
    \item \textit{correlated\_signal\_attack}: CAN messages conveying four wheel speeds, each of which is a two-byte signal on the CAN message, are injected with four different fake wheel speeds. This effectively kills the car—it rolls to a stop and inhibits the driver from effecting acceleration, usually until the car is restarted.
    \item \textit{max\_engine\_coolant\_temp\_attack}:Change the signal segment (one byte) corresponding to the CAN message conveying the temperature of the engine coolant to the maximum value (0xFF). Keep the "engine coolant too high" warning light flashing on the dashboard.
    \item \textit{max\_speedometer\_attack}:Modify the corresponding field (one byte) of the CAN message that transmits the speed signal to the maximum value (0xFF). This will also cause the speed on the dashboard to continue showing the maximum value.
    \item \textit{reverse\_light\_attack}:Modify the corresponding field (one bit) of the CAN message that transmits the status signal of the reversing light, so that the reversing light is continuously on/off.
\end{itemize}
\begin{table*}[ht] % 使用 table* 横跨两列，并用 [ht] 控制位置
\centering
\caption{Performance of ATHENA on the ROAD dataset} % 表格标题
\label{ATHENA_performance}
\begin{tabular}{|c|c|c|c|c|c|c|} % 添加一列 AUC
\hline
\textbf{Method} & \textbf{Attack} & \textbf{Accuracy} & \textbf{Precision} & \textbf{Recall} & \textbf{F1-Score} & \textbf{AUC} \\ \hline
\multirow{8}{*}{\centering \textbf{ATHENA}} 
& \textit{correlated\_signal\_attack} & 0.9776 & 0.9452 & 1.0000 & 0.9718 & 0.98 \\ \cline{2-7}
& \textit{max\_engine\_coolant\_temp\_attack} & 0.9967 & 0.9773 & 1.0000 & 0.9885 & 1.00 \\ \cline{2-7}
& \textit{max\_speedometer\_attack} & 0.9872 & 0.9619 & 0.9800 & 0.9708 & 0.99 \\ \cline{2-7}
& \textit{reverse\_light\_attack} & 0.9920 & 0.9939 & 0.9760 & 0.9849 & 0.99 \\ \cline{2-7}
& \textit{correlated\_signal\_attack\_masquerade} & 0.9167 & 0.9996 & 0.9563 & 0.9166 & - \\ \cline{2-7}
& \textit{max\_engine\_coolant\_temp\_attack\_masquerade} & 0.9821 & 0.7111 & 0.8249 & 0.7138 & - \\ \cline{2-7}
& \textit{max\_speedometer\_attack\_masquerade} & 0.9170 & 0.9293 & 0.9231 & 0.8663 & - \\ \cline{2-7}
& \textit{reverse\_light\_attack\_masquerade} & 0.8304 & 0.7193 & 0.7709 & 0.7115 & - \\ \hline
\end{tabular}
\end{table*}

\begin{figure*}[ht] % 横跨两列的图片
    \centering
    \includegraphics[width=\textwidth]{"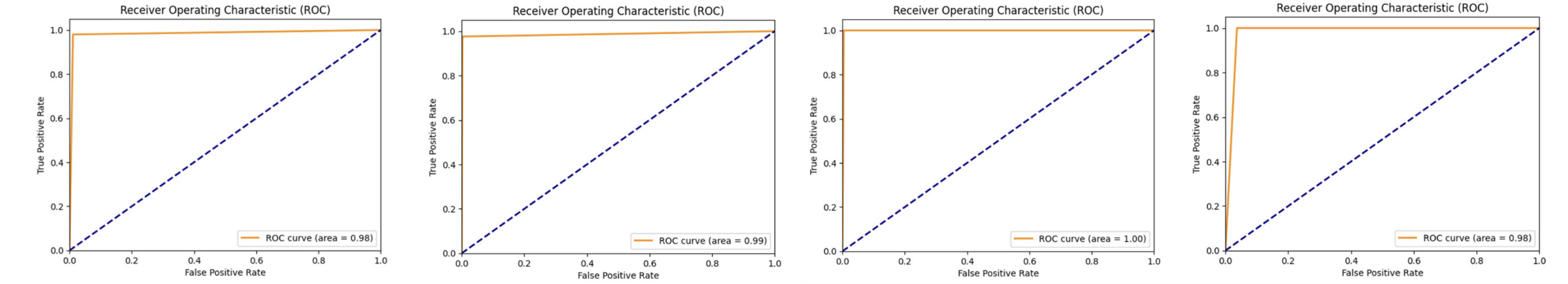"} % 替换为实际图片文件名
    \caption{ROC of ATHENA non-masquerade attack detection. From left to right, they are max\_speedometer\_attack, reverse\_light\_attack, max\_engine\_coolant\_temp\_attack, and correlated\_signal\_attack.}
    \label{fig_8} % 图的引用标签
\end{figure*}

\subsection{Detection Performance}\label{sec5.5}

To evaluate ATHENA's performance, detailed in Table \ref{ATHENA_performance}, we conducted several comparative experiments. The results for accuracy, precision, recall, F1 score, and AUC on other datasets are shown in Table \ref{other_method} (for comparison with state-of-the-art intrusion detection methods, showing optimal performance). Notably, the PCVS module for masquerade attack detection does not use machine learning and directly classifies each sample as an attack or normal, without generating predicted probabilities. Therefore, no ROC curve or AUC is included.

The table shows that ATHENA performs well in most attack scenarios, showing high accuracy, precision, recall, and F1-Score with stable results. It excels in detecting non-masquerade attacks. For masquerade attacks, it achieves an average accuracy of 91.16\% and precision of 83.98\%. These results demonstrate ATHENA's strong detection ability and stability, even with challenging masquerade attacks and unbalanced samples, ensuring effective real-world attack detection.

In comparison, EfficientNet and MobileNet achieve accuracy rates of 96.99\% and 96.37\% on the ROAD dataset, respectively, which are close to ATHENA’s performance. However, their low precision may lead to more false positives in unbalanced scenarios. While both methods perform well in common attack detection, they struggle with masquerade attacks or complex scenarios. LSTM, with a high recall of 95.77\%, shows good detection of positive examples but lacks ATHENA's unique data reconstruction and "normal sending mode" detection, resulting in lower accuracy (67.41\%). It mainly performs well in non-masquerade attack detection but suffers from more false positives overall. CANet shows fair accuracy (94.59\%) but low precision and F1-score, making it less effective than ATHENA in balancing false positives and true positives. RF and MLP perform reasonably well in AUC but lag behind ATHENA in overall performance and stability. G-IDCS TH\_classifier, the first graph-based IDS for IVNs, performs well on specific datasets but poorly on the ROAD dataset, especially under high covert attacks, where its ability to handle masquerade and complex attack scenarios is limited.

\subsection{Hyperparameter Impact on Performance}\label{sec5.6}
\begin{table*}[ht] % 横跨两列
\centering
\caption{Average performance of the other methods on the ROAD dataset} % 表格标题
\label{other_method}
\begin{tabular}{|c|c|c|c|c|c|}
\hline
\textbf{Method} & \textbf{Accuracy} & \textbf{Precision} & \textbf{Recall} & \textbf{F1-Score} & \textbf{AUC} \\ \hline
\textit{EfficientNet} & \textbf{0.9699} & 0.5281 & 0.5232 & 0.5236 & 0.69 \\ \hline
\textit{MobileNet} & 0.9637 & 0.5228 & 0.5084 & 0.5132 & 0.76 \\ \hline
\textit{CANet} & 0.9459 & 0.4111 & 0.3292 & 0.3585 & 0.51 \\ \hline
\textit{G-IDCS TH\_classifier} & 0.7281 & 0.3115 & 0.27 & 0.2811 & 0.33 \\ \hline
\textit{RF} & \textbf{0.9699} & 0.4114 & 0.3429 & 0.3752 & 0.74 \\ \hline
\textit{MLP} & 0.8341 & 0.6970 & 0.4181 & 0.5227 & 0.76 \\ \hline
\textit{ATHENA} & 0.9500 & \textbf{0.9047} & \textbf{0.9289} & \textbf{0.8905} & \textbf{0.99} \\ \hline
\end{tabular}
\end{table*}

%\begin{figure*}[ht] % 横跨两列的图片
%    \centering
%    \includegraphics[width=\textwidth]{"images/9_difference.eps"} % 替换为实际图片文件名
%    \caption{Effect of different hyper-parameters on performance and efficiency of ATHENA.}
%    \label{fig_9} % 图的引用标签
%\end{figure*}
\begin{figure*}[ht]
    \begin{minipage}{0.32\linewidth}
        \centerline{\includegraphics[width=\textwidth]{"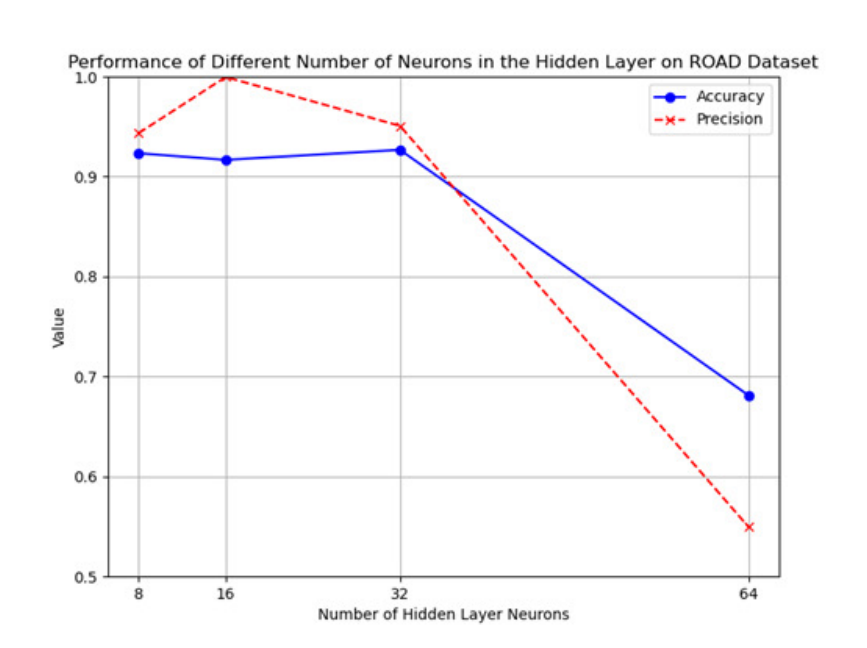"}}
        \centerline{a) Number of Neurons}
    \end{minipage}
    \begin{minipage}{0.32\linewidth}
        \centerline{\includegraphics[width=\textwidth]{"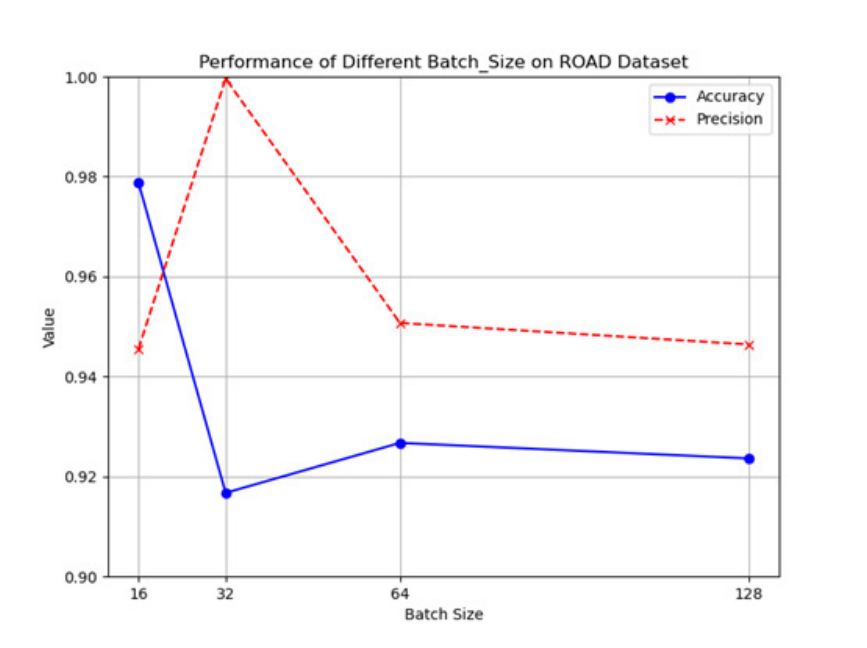"}}
        \centerline{b) Batch Size}
    \end{minipage}
    \begin{minipage}{0.32\linewidth}
        \centerline{\includegraphics[width=\textwidth]{"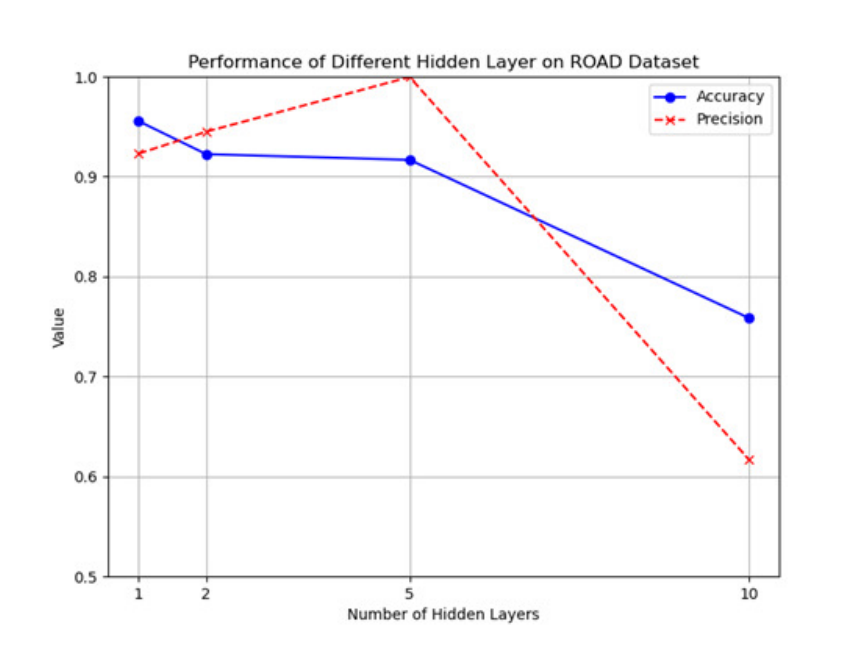"}}
        \centerline{c) Hidden Layers}
    \end{minipage}
    \caption{Comparison of Hyperparameter Tuning on the ROAD Dataset}
    \label{fig_9}
\end{figure*}

\begin{table*}[ht] % 使用 table* 横跨两列，并用 [ht] 控制位置
\centering
\caption{Performance of LSTM without ATHENA assistance on the ROAD dataset} % 表格标题
\label{LSTM_performance}
\begin{tabular}{|c|c|c|c|c|c|} % 删除了最左侧的一列
\hline
\textbf{Attack} & \textbf{Accuracy} & \textbf{Precision} & \textbf{Recall} & \textbf{F1-Score} & \textbf{AUC} \\ \hline
\textit{correlated\_signal\_attack} & 0.3695 & 1.0000 & 0.0005 & 0.0010 & 0.50 \\ \hline
\textit{max\_engine\_coolant\_temp\_attack} & 0.9336 & 0.7018 & 0.9302 & 0.7959 & 0.93 \\ \hline
\textit{max\_speedometer\_attack} & 0.9864 & 0.9634 & 0.9802 & 0.9717 & 0.98 \\ \hline
\textit{reverse\_light\_attack} & 0.9920 & 0.9934 & 0.9755 & 0.9849 & 0.99 \\ \hline
\textit{correlated\_signal\_attack\_masquerade} & 0.3680 & 0.3333 & 0.0005 & 0.0010 & 0.50 \\ \hline
\textit{max\_engine\_coolant\_temp\_attack\_masquerade} & 0.8521 & 0.6191 & 0.3023 & 0.4063 & 0.63 \\ \hline
\textit{max\_speedometer\_attack\_masquerade} & 0.3366 & 0.5234 & 0.9231 & 0.0429 & 0.53 \\ \hline
\textit{reverse\_light\_attack\_masquerade} & 0.3479 & 0.6043 & 0.0464 & 0.0865 & 0.49 \\ \hline
\end{tabular}
\end{table*}

Figure \ref{fig_9} shows how variations in batch size, hidden layer neurons, and hidden layers affect the model’s performance on the ROAD dataset.

From Figure \ref{fig_9} (a), a batch size of 32 achieves the best accuracy and precision. Smaller sizes (16) underperform, while larger ones (128) reduce performance, possibly due to poor generalization or inefficiency. Thus, 32 was chosen to balance performance and efficiency.

Figure \ref{fig_9} (b) shows the effect of hidden layer neurons. With 16 neurons, both accuracy and precision are maximized, while increasing the neurons to 32 and 64 leads to a sharp decline in performance. This suggests overfitting might occur with more neurons, so 16 neurons were chosen for the experiment.

Figure \ref{fig_9} (c) shows the impact of hidden layers on model performance. A network with 5 hidden layers shows optimal accuracy and precision, while fewer layers (1 or 2) and more layers (6 or 10) lead to decreased performance. This suggests that a moderate depth (5 layers) strikes the right balance between model capacity and overfitting, which is why we selected 5 hidden layers for my experiment.

Thus, based on these results, the chosen hyperparameters—batch size of 32, 16 neurons per hidden layer, and 5 hidden layers—were identified as the best configuration for optimizing model performance on the ROAD dataset.

\subsection{Ablation experiment}\label{sec5.6}

In this section, we discuss the effectiveness of important individual components in the ATHENA framework, and specifically, we contrast the difference in performance between LSTM without the assistance of the ATHENA framework and ATHENA-LSTM. As shown in Table \ref{LSTM_performance}, the ability of LSTM to detect Timing Transparent attacks is significantly reduced, and the results become unbalanced after the absence of the process of data cloud preprocessing and delivery of ATHENA framework. In the detection of impersonation attacks, the Recall and F1-score values are extremely low, which shows that there is a serious performance bottleneck in the detection of impersonation attacks by LSTM, which needs to rely on other components in the ATHENA framework to break through. Therefore, compared with the traditional LSTM, ATHENA-LSTM has great changes and careful design according to the uniqueness of the use environment.

\section{conclusion}\label{section6}
In this paper, we introduced the ATHENA framework, a novel intrusion detection system for in-vehicle networks that leverages both vehicle system physics and a vehicle-cloud integrated architecture. The framework includes two core methods: PCVS, which combines first-principle physics with data-driven rule mining to address masquerade attacks, and ATHENA-LSTM, which captures long-term dependencies in CAN message timing to detect timing-based attacks. By offloading computationally intensive tasks to the cloud and lightweight detection tasks to the vehicle terminal, ATHENA optimizes resource allocation and balances high computational power with low resource consumption.

Experimental results confirm ATHENA's effectiveness and scalability in real-world vehicle environments, demonstrating its ability to handle complex attacks with real-time response. This work introduces a groundbreaking approach to intrusion detection in the Internet of Vehicles, advancing in-vehicle network security. Future efforts will focus on improving robustness and expanding its applicability to more vehicle models and network configurations.

\bibliographystyle{IEEEtran}

%\bf{If you include a photo:}
\vspace{-40pt}
\begin{IEEEbiography}[{\includegraphics[width=1in,height=1.25in,clip,keepaspectratio]{"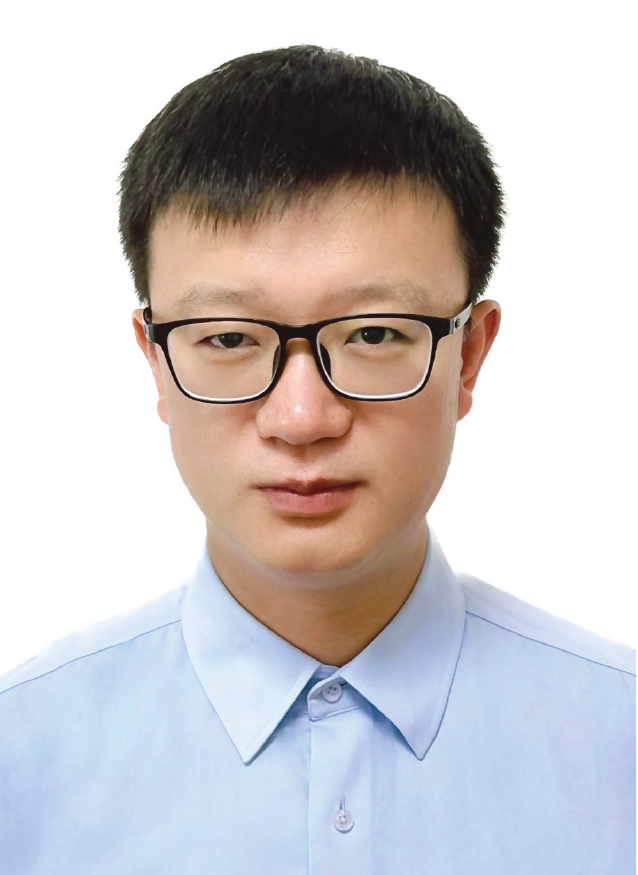"}}]{Kai Wang} (Member, IEEE) 
is a full Professor with School of Computer Science and Technology, Harbin Institute of Technology (HIT), Weihai, China. His research interests include in-vehicle network security, advanced persistent threat (APT) detection, and trustworthy machine learning. He has published more than 40 papers in prestigious international journals, including IEEE TITS, ACM TOIT, ACM TIST, etc. He received the Ph.D. degree in Communication and Information Systems from Beijing Jiaotong University, China, in 2014. He is a Member of the IEEE and ACM, and a Senior Member of the China Computer Federation (CCF). Contact him at email: dr.wangkai@hit.edu.cn.
\end{IEEEbiography}

\vspace{-30pt}\begin{IEEEbiography}[{\includegraphics[width=1in,height=1.25in,clip,keepaspectratio]{"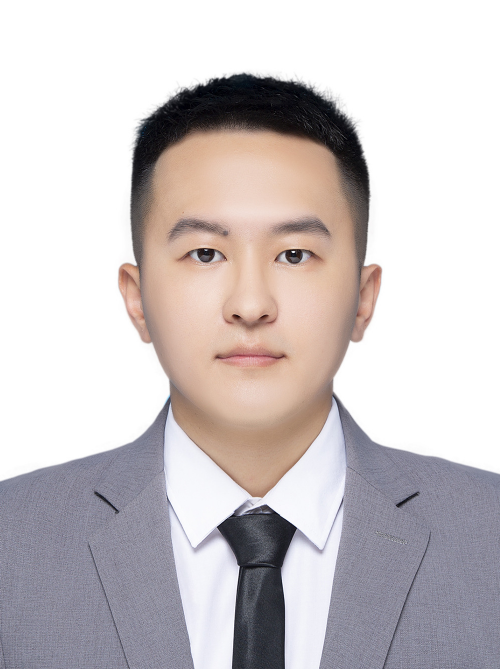"}}]{Zhen Sun}
is a master's student majoring with School of Computer Science and Technology, Harbin Institute of Technology (HIT), Weihai, China. His research interests include intelligent and efficient in-vehicle intrusion detection models. He received the B.S. degree in Information and Computing Science from Civil Aviation University of China. Contact him at email: 23s130402@stu.hit.edu.cn.
%received the B.S. degree in Information and Computing Science from Civil Aviation University of China, China. He is currently pursuing the master's degree in computer technology with the Harbin Institute of Technology (HIT), China. Her research interests include intelligent and efficient in-vehicle intrusion detection models.
%\vspace{-5pt}
\end{IEEEbiography}

\vspace{-30pt}
\begin{IEEEbiography}[{\includegraphics[width=1in,height=1.25in,clip,keepaspectratio]{"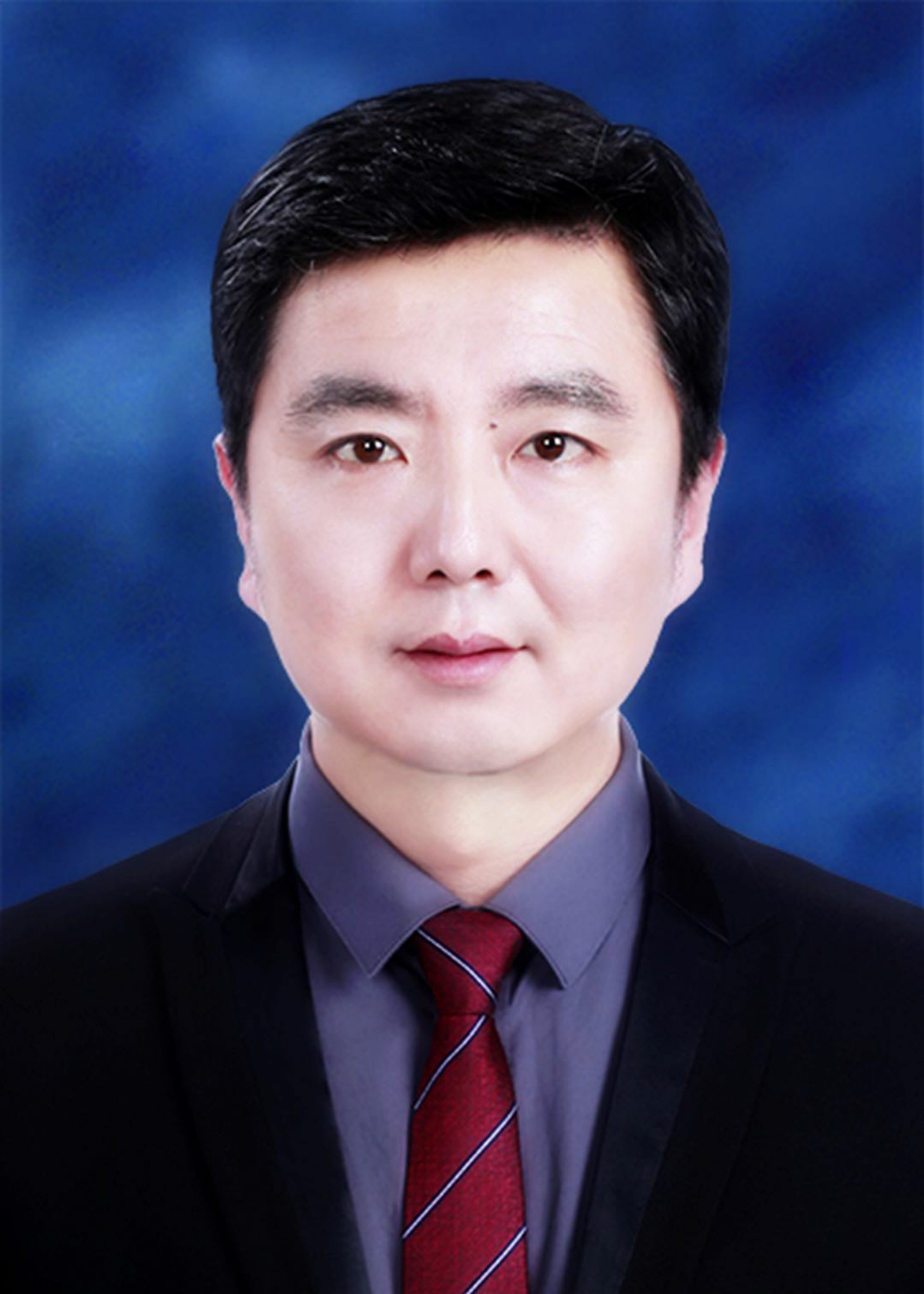"}}]{Bailing Wang} (Member, IEEE) 
is a full Professor with School of Computer Science and Technology, Harbin Institute of Technology (HIT), Weihai, China. His research interests include information content security and industrial control network security. He has published more than 80 papers in prestigious international journals and been selected for the China national talent plan. He received the Ph.D. degree in Computer Architecture from Harbin Institute of Technology (HIT), China, in 2006. Contact him at email: wbl@hit.edu.cn.
%received the Ph.D. degree from the School of Computer Science and Technology, Harbin Institute of Technology (HIT), China, in 2006. He is currently a Professor with the Faculty of Computing, HIT. He has published more than 80 papers in prestigious international journals and conferences, and has been selected for the China national talent plan. His research interests include information content security, industrial control network security, and V2X security.
\end{IEEEbiography}

\vspace{-30pt}
\begin{IEEEbiography}[{\includegraphics[width=1in,height=1.25in,clip,keepaspectratio]{"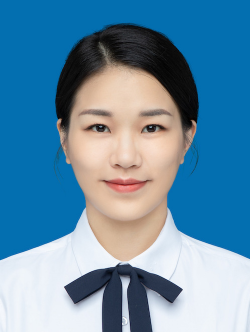"}}]{Qilin Fan} (Member, IEEE) 
is an Associate Professor in the School of Big Data and Software Engineering, Chongqing University, Chongqing, China. Her research interests include federated learning, mobile edge computing, and machine learning. She received the B.E. degree in the College of Software Engineering, Sichuan University, Chengdu, China, in 2011, and the Ph.D. degree from the Department of Computer Science and Technology, Tsinghua University, Beijing, China, in 2017. She is a member of the IEEE and the China Computer Federation (CCF). Contact her at email: fanqilin@cqu.edu.cn.
\end{IEEEbiography}

\vspace{-30pt}
\begin{IEEEbiography}[{\includegraphics[width=1in,height=1.25in,clip,keepaspectratio]{"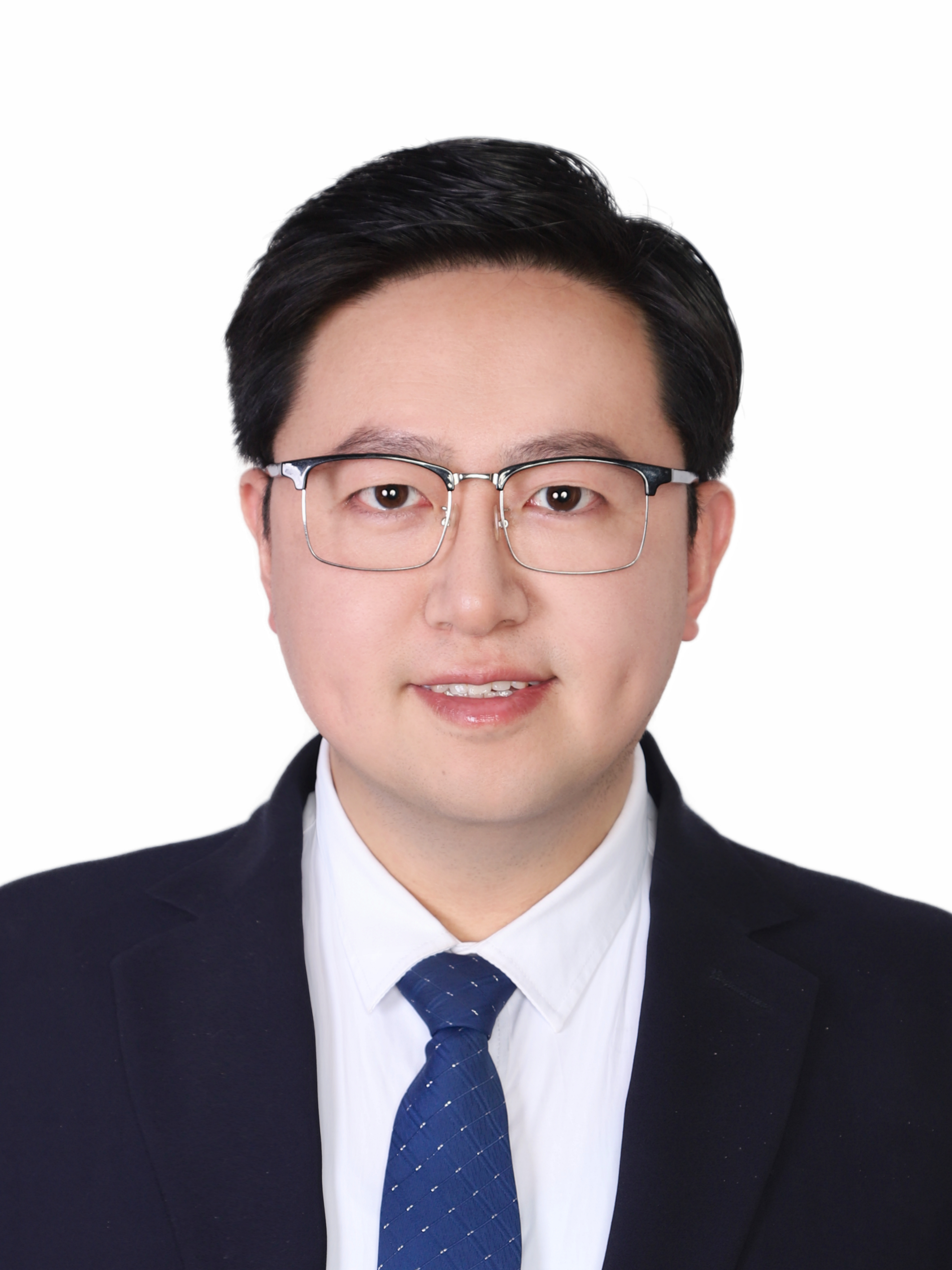"}}]{Ming Li} (Member, IEEE) 
is the Vice General Manager and CTO at Shandong Inspur Database Technology Co., Ltd. His research interests include HTAP cloud-native high-performance databases, big data analytics systems, and edge computing. He has published over 10 papers and books also participated in several national standards. He received the Ph.D. degree in Communication Networks from Hamburg University of Technology, Germany, in 2017. He is an Executive Director of the China Industry-University-Research Institute Collaboration Association (CIUR). Contact him at email: liming2017@inspur.com. 
%obtained his B.S. degree from Shandong University, M.Sc. degree from Ulm University, and Ph.D. degree from Hamburg University of Technology. He is currently the Vice General Manager and CTO at Shandong Inspur Database Technology Co., Ltd. Before joining the Inspur Group, he worked as a senior engineer at Intel Corporation. He has published over 10 papers and books, including IEEE GLOBOCOM, ICC, ITC, WCNC, VTC, etc., and has also participated in several national standards. His current research interests include HTAP cloud-native high-performance databases, big data analytics systems, and edge computing. He serves as an Executive Director of the China Industry-University-Research Institute Collaboration Association (CIUR) and is the Director of the Jinan Key Laboratory of Distributed Databases.
\end{IEEEbiography}

\vspace{-30pt}
\begin{IEEEbiography}[{\includegraphics[width=1in,height=1.25in,clip,keepaspectratio]{"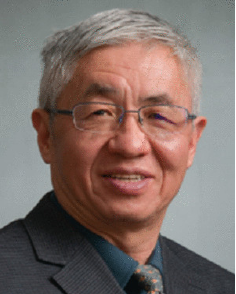"}}]{Hongke Zhang}(Fellow, IEEE) 
is a full Professor with School of Electronic and Information Engineering, Beijing Jiaotong University, Beijing, China, and also an Academician of China Engineering Academy. His research interests include architecture and protocol design for the future Internet and specialized networks. He received the Ph.D. degree in Communication and Information Systems from the University of Electronic Science and Technology of China, Chengdu, China, in 1992. He directs the National Engineering Center of China on Mobile Specialized Network, and serves as an Associate Editor for the IEEE Transactions on Network and Service Management and IEEE Internet of Things Journal. Contact him at email: hkzhang@bjtu.edu.cn.
%is a Professor with the School of Electronic and Information Engineering, Beijing Jiaotong University, Beijing, China, where he currently directs the National Engineering Center of China on Mobile Specialized Network. He received the Ph.D. degree in communication and information system from the University of Electronic Science and Technology of China, Chengdu, China, in 1992. His current research interests include architecture and protocol design for the future Internet and specialized networks. He currently serves as an Associate Editor for the IEEE Transactions on Network and Service Management and IEEE Internet of Things Journal. He is an Academician of China Engineering Academy.
\end{IEEEbiography}

\end{document}